\pdfoutput=1

\documentclass[11pt]{article}

\usepackage[final]{ACL2023}

\usepackage{times}
\usepackage{latexsym}

\usepackage[T1]{fontenc}

\usepackage[utf8]{inputenc}

\usepackage{microtype}

\usepackage{inconsolata}

\usepackage{multirow}
\usepackage{array,makecell,booktabs}
\newcolumntype{M}[1]{>{\centering\arraybackslash}m{#1}}
\usepackage{wrapfig}

\usepackage{graphicx}
\usepackage{float}
\usepackage{overpic}
\usepackage{wrapfig}
\usepackage{caption}
\usepackage{subfigure}

\usepackage{amsmath}

\title{FluentSpeech: Stutter-Oriented Automatic Speech Editing \\ with Context-Aware Diffusion Models}

\author{Ziyue Jiang\thanks{~~Equal contribution.} \\
  Zhejiang University\\
  \texttt{ziyuejiang@zju.edu.cn} \\\And
  Qian Yang\footnotemark[1] \\
  Zhejiang University \\
  \texttt{qyang1021@zju.edu.cn} \\\And
  Jialong Zuo \\
  Zhejiang University \\
  \texttt{jialongzuo@gmail.com} \\\AND
  Zhenhui Ye \\
  Zhejiang University \\
  \texttt{zhenhuiye@zju.edu.cn} \\\And
  Rongjie Huang\\
  Zhejiang University \\
  \texttt{rongjiehuang@zju.edu.cn} \\\And
  Yi Ren \\
  Bytedance AI Lab \\
  \texttt{ren.yi@bytedance.com} \\\AND
  Zhou Zhao\thanks{~~Corresponding author.} \\
  Zhejiang University \\
  \texttt{zhaozhou@zju.edu.cn} \\}

\begin{document}
\maketitle

\begin{abstract}
Stutter removal is an essential scenario in the field of speech editing. However, when the speech recording contains stutters, the existing text-based speech editing approaches still suffer from: 1) the over-smoothing problem in the edited speech; 2) lack of robustness due to the noise introduced by stutter; 3) to remove the stutters, users are required to determine the edited region manually. To tackle the challenges in stutter removal, we propose FluentSpeech, a stutter-oriented automatic speech editing model. Specifically, 1) we propose a context-aware diffusion model that iteratively refines the modified mel-spectrogram with the guidance of context features; 2) we introduce a stutter predictor module to inject the stutter information into the hidden sequence; 3) we also propose a stutter-oriented automatic speech editing (SASE) dataset that contains spontaneous speech recordings with time-aligned stutter labels to train the automatic stutter localization model. Experimental results on VCTK and LibriTTS datasets demonstrate that our model achieves state-of-the-art performance on speech editing. Further experiments on our SASE dataset show that FluentSpeech can effectively improve the fluency of stuttering speech in terms of objective and subjective metrics. Code and audio samples can be found at \url{https://github.com/Zain-Jiang/Speech-Editing-Toolkit}.









\end{abstract}

\section{Introduction}
Recently, text-based speech editing~\citep{jin2017voco,jin2018speech,morrison2021context,tan2021editspeech,tae2021editts,wang2022context,bai20223} has made rapid progress, and stutter removal is a critical sub-task in speech editing. There are various application scenarios for stutter removal, like short-form videos, movies, podcasts, YouTube videos, and online lectures, since it provides great convenience for media producers.

Previous speech editing systems~\cite{jin2017voco,jin2018speech} successfully enable the user to edit the speech recording through operations in the text transcript. Some neural text-to-speech (TTS) based methods~\citep{tan2021editspeech,tae2021editts} achieve smooth transition at the boundaries of the edited region. And most recently, the mask prediction based methods~\citep{wang2022context,bai20223} learn better contextual information from the input mel-spectrogram and outperform previous approaches at speech quality and prosody modeling. However, the existing approaches only aim at modifying reading-style speeches, while removing stutters from spontaneous speeches remains a considerable challenge.

When applied to the stutter removal task, previous efforts are still subject to the following limitations: 1) the generated mel-spectrogram is usually blurry and lacks frequency bin-wise details, resulting in unnatural sounds in the boundaries of the modified region; 2) when the speech recording is full of stutters, the edited speech is usually not robust due to the noise introduced by the discrepancy between text and stuttering speech content; 3) the stutter region should be manually determined one by one, which is costly and laborious for media producers.

To tackle these challenges, we propose FluentSpeech, the first generative model to solve the stutter removal task, which automatically detects the stutter regions, removes them, and generates fluent speech with natural details. Specifically,

\begin{itemize}

\item Non-probabilistic models tend to generate over-smooth mel-spectrograms~\citep{huang2022prodiff,popov2021grad}, while probabilistic models (e.g., GAN and diffusion) generate mel-spectrograms with richer frequency details and natural sounds. Based on this observation, we adopt a context-aware diffusion model that utilizes rich contextual information to guide the diffusion and reverse processes, which helps FluentSpeech to generate high-quality and expressive results.

\item To improve the robustness against stuttering speeches, we introduce a conditional stutter predictor that localizes the stutter region and injects the stutter information into the frame-level hidden sequence to reduce the discrepancy between text and stuttering speech. Moreover, the predicted stutter region can be utilized as the mask for automatic stutter removal.


\item We propose a novel dataset called the stutter-oriented automatic speech editing (SASE) dataset, which contains spontaneous speech recordings with time-aligned stutter labels for automatic stutter removal.

\end{itemize}

Experiments on the VCTK~\citep{yamagishi2019vctk} and LibriTTS~\citep{zen2019libritts} dataset show that FluentSpeech outperforms state-of-the-art models on speech editing towards reading-style speech with fewer model parameters. And in the experiments on our newly collected SASE dataset, FluentSpeech enjoys much robustness against stuttering speech and demonstrates the ability to improve the fluency of stuttering speech significantly. The main contributions of this work can be summarized as follows:

\begin{itemize}

\item We analyze the characteristics of different speech editing approaches (e.g., algorithm, architecture, alignment learning approaches, etc.) and propose a context-aware diffusion probabilistic model that achieves state-of-the-art performance on speech editing.

\item We propose a stutter predictor module to improve the robustness against the stuttering speech and localize the stutter region. The stutter predictor can also control the stutter representations by removing the stutters from the spontaneous speech to improve its fluency, which solves the automatic stutter removal task for the first time.

\item We contribute a novel SASE dataset which contains 40 hours of spontaneous speech crawled from online lectures or open courses given by 46 speakers. We will publish our model and dataset as the benchmark for the evaluation of future SASE algorithms.

\end{itemize}

\section{Background}
In this section, we describe the background of speech editing and the basic knowledge of diffusion model. We also review the existing applications of diffusion model in speech tasks and analyze their advantages and disadvantages.
\subsection{Speech Editing}
Conventional speech editing methods~\citep{derry2012pc,whittaker2004semantic} provide users with interfaces for cut, copy, paste, volume adjustment, time-stretching, pitch bending, de-noising, etc. Then text-based speech editing systems~\citep{jin2017voco,jin2018speech} allow the editor to perform select, cut, and paste operations in the text transcript of the speech and apply the changes to the waveform accordingly. However, they mainly face two problems. One is that the edited speech often sounds unnatural because the edited region does not match the prosody of the speech context. (e.g., mismatches in intonation, stress, or rhythm)~\citep{jin2017voco}. Another is that the interfaces do not support the ability to synthesize new words not appearing in the transcript~\citep{morrison2021context}. There are a series of studies on these problems. ~\citet{jin2017voco} propose to insert a synthesized audio clip using a combination of the text-to-speech model and voice conversion model~\citep{sun2016phonetic}, which leads to unnatural prosody near the boundaries of the edited regions. ~\citet{tan2021editspeech} use neural TTS model with auto-regressive partial inference to maintain a coherent prosody and speaking style. Most recently, the mask prediction based methods~\citep{wang2022context,bai20223} can capture more contextual information from the input mel-spectrogram. ~\citet{wang2022context} propose to learn the relation between text and audio through cross-attention but suffer from the extremely slow convergence rate. ~\citet{bai20223} introduce the alignment embedding into the Conformer-based~\citep{gulati2020conformer,guo2021recent} backbone to improve the speech quality. However, previous methods only focus on the modification of reading-style speeches, which is not stutter-oriented.

\begin{figure*}[ht]
    \centering
    \small
    \subfigure[FluentSpeech]{
        \includegraphics[width=.235\linewidth]{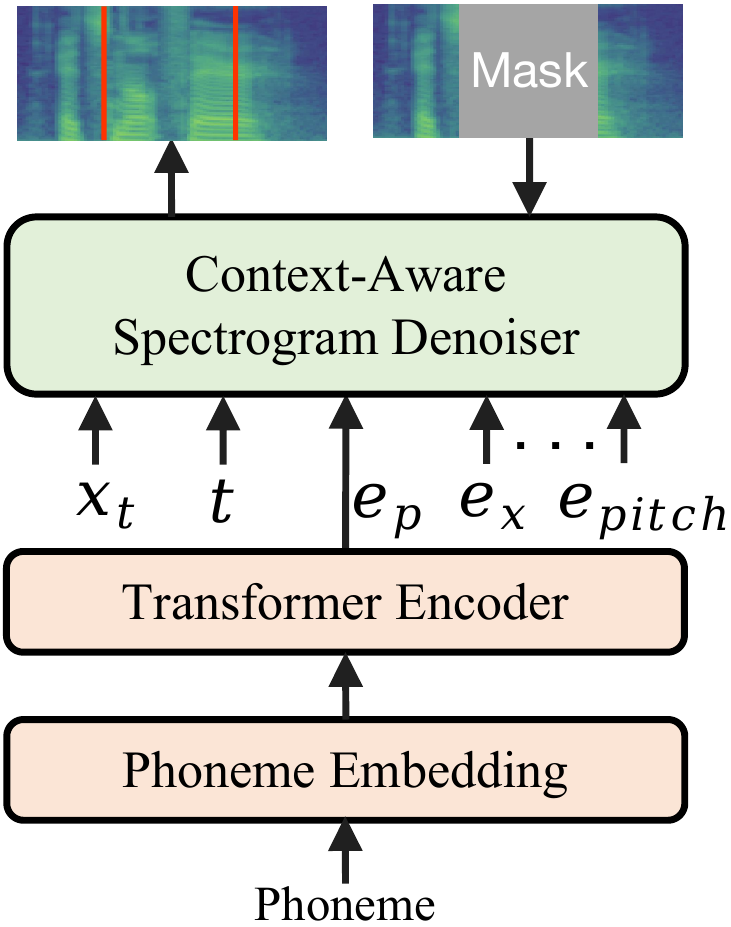}
    }
    \subfigure[Reverse Diffusion]{
        \includegraphics[width=.172\linewidth]{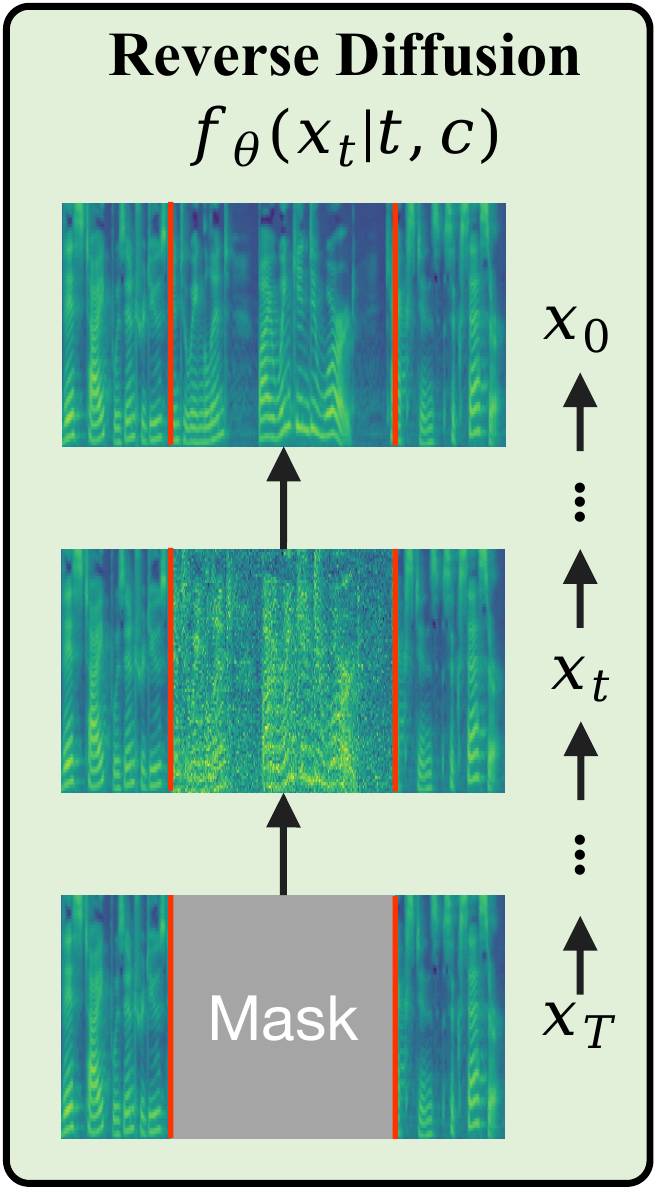}
    }
    \subfigure[Context-Aware Spectrogram Denoiser]{
        \hspace{3mm}
        \includegraphics[width=.50\linewidth]{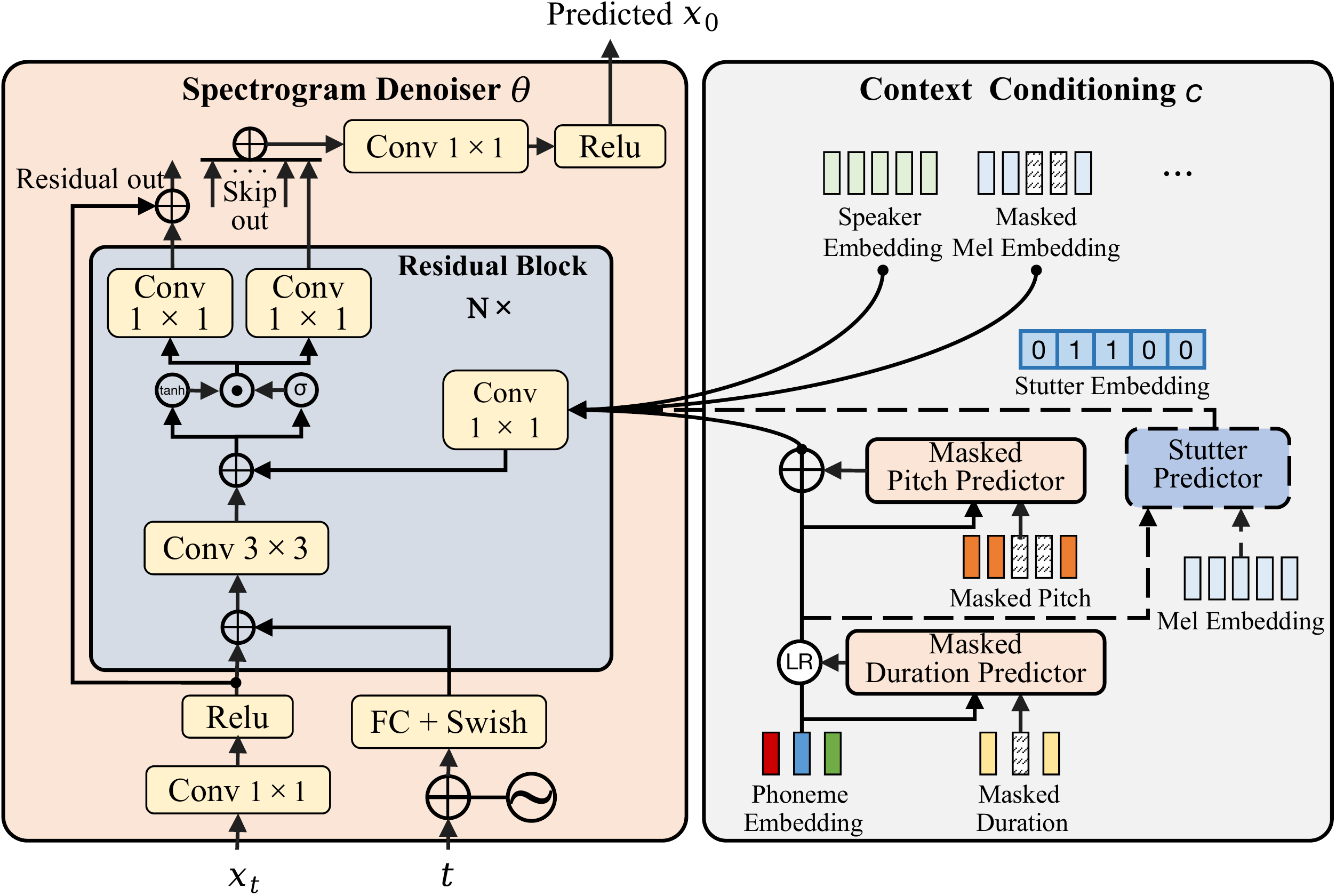}
        \label{arch_c}
    }
\caption{The overall architecture for FluentSpeech. In subfigure (c), the spectrogram denoiser $\theta$ takes noisy spectrogram $x_{t}$ as input and computes $f_{\theta}(x_{t}|t,c)$ conditioned on diffusion time index $t$ and context information $c$. The sinusoidal-like symbol, FC, Swish, and $\bullet$ denote the positional encoding, fully-connected layer, swish activation function~\citep{ramachandran2017searching}, and element-wise multiple operation. $LR$ denotes the Length Regulator proposed in FastSpeech~\citep{ren2019fastspeech}. $N$ is the number of residual blocks. The dashed black line denotes that the operation is only executed when the dataset contains spontaneous speeches.}
\label{main_architecture}
\end{figure*}

\subsection{Diffusion Model}
\paragraph{Basic knowledge of diffusion model} Denoising diffusion probabilistic models (DDPMs) have achieved state-of-the-art performances in both image and audio synthesis~\citep{dhariwal2021diffusion,kong2020diffwave,huang2022prodiff}. DDPMs~\citep{ho2020denoising,dhariwal2021diffusion} are designed to learn a data distribution $p(x)$ by gradually denoising a normally distributed variable through the reverse process of a fixed Markov Chain of length $T$. Denote $x_{t}$ as a noisy version of the clean input $x_{0}$. DDPMs choose to parameterize the denoising model $\theta$ through directly predicting $\boldsymbol{\epsilon}$ with a neural network $\boldsymbol{\epsilon}_{\theta}$.  The corresponding objective can be simplified to: \begin{equation}
\small
\mathcal{L}_\theta^{\mathrm{Grad}}=\left\|\boldsymbol{\epsilon}_\theta\left(\alpha_t \boldsymbol{x}_0+\sqrt{1-\alpha_t^2} \boldsymbol{\epsilon}\right)-\boldsymbol{\epsilon}\right\|_2^2, \boldsymbol{\epsilon} \sim \mathcal{N}(0, \mathrm{I}),
\end{equation}
with $t$ uniformly sampled from $\{1, ... , T\}$.
\paragraph{Applications of diffusion model in speech tasks} Applications of diffusion model in speech tasks mainly lie in speech synthesis. Diff-TTS~\citep{jeong2021diff}, Grad-TTS~\citep{popov2021grad}, and DiffSpeech~\citep{liu2021diffsinger} are gradient-based models with score-matching objectives to generate high-quality speeches, which require hundreds of iterations with small $\beta_{t}$ to guarantee high sample quality. Most recently, ProDiff~\citep{huang2022prodiff} parameterize the denoising model by directly predicting clean data and avoids significant perceptual quality degradation when reducing reverse iterations. In the field of speech editing, ~\citet{tae2021editts} propose a diffusion model that requires a pre-trained TTS model to synthesize the target audio and eliminate the artifacts of concatenation by a score-based manipulation algorithm, which is not text-based speech editing.

\section{FluentSpeech}
This section presents our proposed FluentSpeech, a stutter-oriented automatic speech editing model that solves the stutter removal task. We firstly overview the motivation and the architecture of FluentSpeech. Secondly, we describe the detailed designs of alignment modeling, context-aware spectrogram denoiser, and stutter predictor. Finally, we describe the training objectives of FluentSpeech, following with the illustration of training and inference procedures.

\subsection{Model Overview}
The overall model architecture of FluentSpeech is shown in Figure~\ref{main_architecture}. FluentSpeech consists of a linguistic encoder and a context-aware spectrogram denoiser. Denote the phoneme sequence of the transcription as $\boldsymbol{p} = (p_1, \ldots, p_{|p|})$ and the acoustic feature sequence as $\boldsymbol{x} = (x_1, \ldots, x_{|x|})$. $\boldsymbol{x}$ can be the spectrogram or mel-spectrogram of the speech audio, and each $x_i$ represents the speech feature of frame $i$. The Transformer-based~\citep{vaswani2017attention} linguistic encoder converts $\boldsymbol{p}$ into the text hidden sequence $\boldsymbol{e}_{p}$. Denote $\hat{\boldsymbol{x}} = Mask(\boldsymbol{x, \lambda})$ as the masked acoustic feature sequence, where $Mask(\cdot)$ replaces several random spans of $x$ by the probability of $\lambda$ with the same number of a random initialized masking vector. Then, the context-aware spectrogram denoiser $\theta$ aggregates phoneme embedding $\boldsymbol{e}_{p}$ and other features like acoustic embedding $\boldsymbol{e}_{\boldsymbol{x}}$, pitch embedding $\boldsymbol{e}_{pitch}$ as the condition $c$ to guide the reverse process of the diffusion model $f_\theta\left(x_t \mid t, c\right)$.



\subsection{Alignment Modeling} Due to the modality gap between text and speech, alignment modeling is essential in text-based speech editing. There are three types of approaches to model the monotonous alignment between text and speech: 1) cross-attention, \citet{wang2022context} propose to learn the alignment information with the cross-attention module in the transformer decoder, which suffers from the slow convergence rate and is usually not robust; 2) alignment embedding, ~\citet{bai20223} introduce the alignment embedding from external alignment tools into the self-attention based architecture to guide the alignment modeling; 3) length regulator~\citep{ren2019fastspeech,tan2021editspeech}, the length regulator expand text embedding into frame-level embedding according to the phoneme duration predicted by the duration predictor~\citep{ren2019fastspeech,tan2021editspeech}, which ensures hard alignments and is more robust than the above two methods. However, the duration predictor in~\citet{tan2021editspeech} does not consider the existing context duration. It only predicts the duration of the entire sentence from text representations and applies the duration of the edited words to the masked region, which results in unnatural prosody. Therefore, in FluentSpeech, we train the duration predictor with the mask prediction procedure to achieve the fluent duration transition at the edited region, which is called the masked duration predictor.

\subsection{Context-Aware Spectrogram Denoiser}
\paragraph{Context Conditioning} As shown in Figure~\ref{arch_c}, in the context conditioning module, we adopt frame-level text embedding $\boldsymbol{e}_{t}$, acoustic feature sequence $\boldsymbol{x}$, masked acoustic feature sequence $\hat{\boldsymbol{x}}$, speaker embedding $\boldsymbol{e}_{spk}$, pitch embedding $\boldsymbol{e}_{pitch}$, and stutter embedding $\boldsymbol{e}_{stutter}$ as the condition for our spectrogram denoiser. The phoneme embedding $\boldsymbol{e}_{p}$ is first expanded into frame-level text embedding $\boldsymbol{e}_{t}$ by the length regulator with the duration information from the masked duration predictor. We add $\boldsymbol{e}_{t}$ to the context condition $c$. We also extract the speaker embeddings $\boldsymbol{e}_{spk}$ from audio samples using open-source voice encoder\footnote{\url{https://github.com/resemble-ai/Resemblyzer}} and feed them into the context condition $c$ following the common practice~\citep{min2021meta,huang2022prodiff,tan2021editspeech}. Then we adopt a nonlinear feed-forward acoustic encoder to transform the speech feature $\boldsymbol{x}$ and $\hat{\boldsymbol{x}}$ into the acoustic embeddings $\boldsymbol{e}_{\boldsymbol{x}}$ and $\boldsymbol{e}_{\hat{\boldsymbol{x}}}$ following ~\citet{bai20223}. The masked acoustic embedding $\boldsymbol{e}_{\hat{\boldsymbol{x}}}$ is also added to the condition to provide more contextual information for mel-spectrogram reconstruction. Moreover, the masked pitch predictor utilizes $\boldsymbol{e}_{t}$ and the masked pitch embedding $\boldsymbol{\hat{e}}_{pitch}$ to predict the pitch $F_0$ of each frame in the edited region. We further convert it into the pitch embedding vector and add it to the context condition $c$. To promote the natural transition at the edited boundaries, we train the duration predictor and pitch predictor with the mask prediction procedure: 
\begin{align}
    \mathcal{L}_p&=\|p- g_{p}(\boldsymbol{e}_{t}, \boldsymbol{\hat{e}}_{pitch}) \|_2^2 \ , \\
    \mathcal{L}_{d}&=\|d- g_{d}(\boldsymbol{e}_{d}, \boldsymbol{\hat{e}}_{dur}) \|_2^2
\end{align}
where we use $d$ and $p$ to denote the target duration and pitch respectively, and use $g_{d}$ and $g_{p}$ to denote the corresponding duration predictor and pitch predictor, which share the same architecture of 1D convolution with ReLU activation and layer normalization. The loss weights are all set to 0.1 and the reconstruction losses are also added to train the linguistic encoder.

\paragraph{Spectrogram Denoiser} Following ~\citet{liu2021diffsinger,huang2022prodiff}, we adopt a non-causal WaveNet~\citep{oord2016wavenet} architecture to be our spectrogram denoiser. The decoder comprises a 1x1 convolution layer and $N$ convolution blocks with residual connections to project the input hidden sequence with 256 channels. For any step $t$, we use the cosine schedule $\beta_{t} = cos(0.5 \pi t)$. Different from the aforementioned diffusion models that require hundreds of steps with small $\beta_{t}$ to estimate the gradient for data density, we choose to parameterize the denoising model by directly predicting the clean data $x_{0}$ following recent researches in image generation and TTS literature~\citep{salimans2021progressive,liu2022diffgan,huang2022prodiff} to significantly accelerate sampling from a complex distribution. Specifically, in the generator-based diffusion models, $p_{\theta}(x_{0}|x_{t})$ is the implicit distribution imposed by the neural network $f_{\theta}(x_{t},t)$ that outputs $x_{0}$ given $x_{t}$. And then $x_{t-1}$ is sampled using the posterior distribution $q(x_{t-1}|x_{t},x_{0})$ given $x_{t}$ and the predicted $x_{0}$. The training loss is defined as the mean absolute error (MAE) in the data $x$ space:
\begin{equation}
\small
\mathcal{L}_{\theta}^{MAE}=\left\|\boldsymbol{x}_\theta\left(\alpha_t \boldsymbol{x}_0+\sqrt{1-\alpha_t^2} \boldsymbol{\epsilon}\right)-\boldsymbol{x}_0\right\|, \boldsymbol{\epsilon} \sim \mathcal{N}(0, \boldsymbol{I})\ ,
\end{equation}
and efficient training is optimizing a random $t$ term with stochastic gradient descent. Inspired by ~\citep{ren2022revisiting}, we also adopt structural similarity index (SSIM) loss $\mathcal{L}_{\theta}^{SSIM}$ in training to capture structural information in mel-spectrogram and improve the perceptual quality:
\begin{equation}
\small
\mathcal{L}_{\theta}^{\mathrm{SSIM}}=1-\operatorname{SSIM}\left(\boldsymbol{x}_\theta\left(\alpha_t \boldsymbol{x}_0+\sqrt{1-\alpha_t^2} \boldsymbol{\epsilon}\right), \hat{x}_0\right)\ .
\end{equation}
The loss weights are all set to 0.5. Since the capability of our spectrogram denoiser is powerful enough, we do not adopt the convolutional Post-Net to refine the predicted spectrogram like previous works~\citep{wang2022context,bai20223}.

\begin{figure}[ht]
    \centering
    \small
    \includegraphics[width=.98\linewidth]{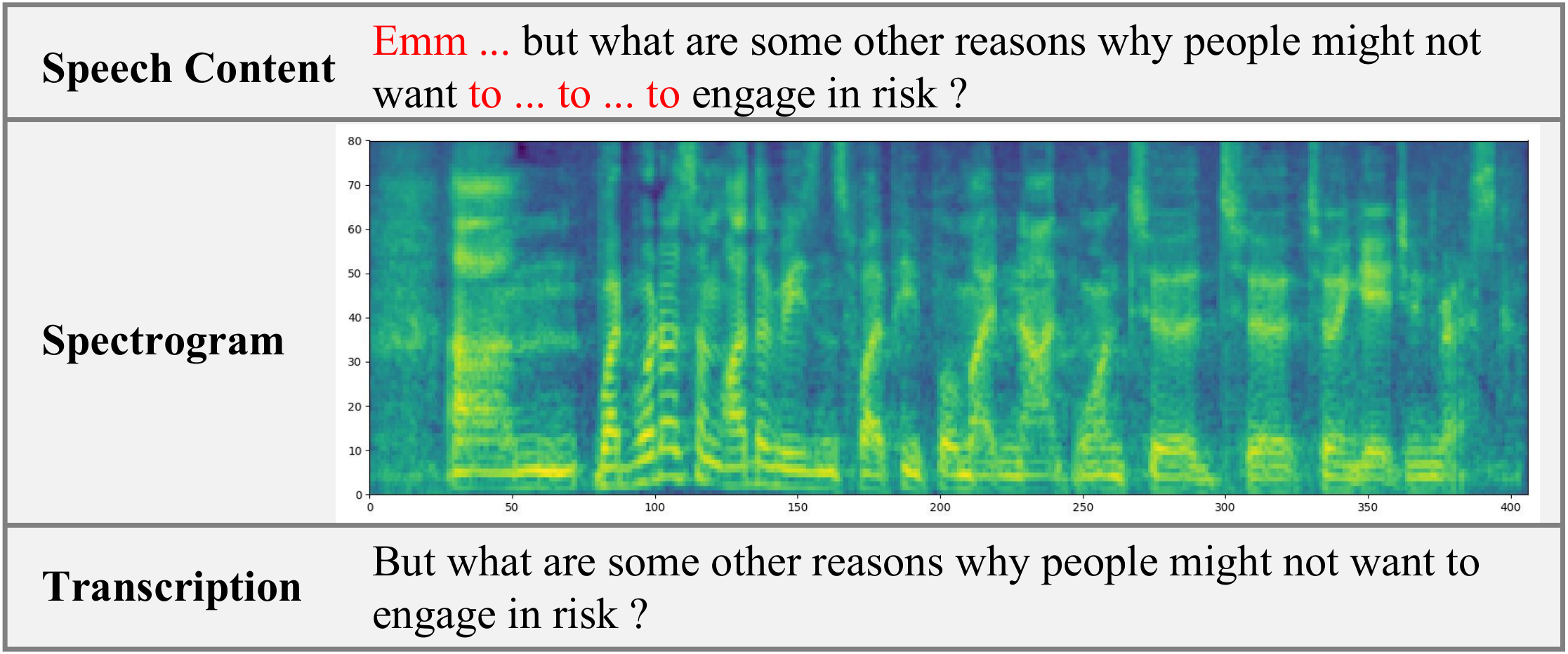}
\caption{The illustration of the discrepancy between the given transcription and stuttering speech content.}
\label{stutter_demonstration}
\end{figure}

\subsection{Stutter Predictor} 
The stutter predictor is introduced only when the speech corpus contains stuttering recordings. The stutters in the speech content will introduce noise to the training pipeline due to the noise introduced by the information gap between text and stuttering speech content. As shown in Figure~\ref{stutter_demonstration}, the stuttering word ``to'' in the speech content makes the speech editing model learn unintentional sounds in the pronunciation of the word ``to''. Therefore, we introduce the stutter embedding into the text hidden sequence to disentangle the stutter-related gradients from the speech content, which significantly improves the pronunciation robustness of our FluentSpeech.

Let $s = (s_1, \ldots, s_{s} )$ be a time-aligned stutter label that defines the stutter regions in the corresponding spontaneous speech, where $s_i \in \{0,1\}$ (0 for normal and 1 for stutter) for each frame (See Appendix~\ref{appendix: sase dataset} for further details about the stutter label in our SASE dataset). In training, we take the ground-truth value of the stutter label as input into the hidden sequence to predict the target speech. At the same time, we use the ground-truth labels as targets to train the stutter predictor, which is used in inference to localize the stutter region in target speech.

The stutter predictor consists of 1) a 4-layer 1D conditional convolutional network with ReLU activation, each followed by the layer normalization and the dropout layer; 2) an extra linear layer and a softmax layer to predict the probability of stutter tag. As shown in Figure~\ref{arch_c}, we propose a text-guided stutter predictor module, which takes frame-level text embedding $\boldsymbol{e}_{t}$ and mel-spectrogram embedding $\boldsymbol{e}_{\boldsymbol{x}}$ as input and seeks to locate the text-irrelevant stutter regions. The main objective function for stutter prediction is the binary cross-entropy loss $\mathcal{L}_{BCE}$. The Focal loss~\citep{lin2017focal} $\mathcal{L}_{Focal}$ is also introduced since the misclassification of fluent regions is tolerable and we want the stuttering regions to be accurately classified. The $\alpha_0$, $\alpha_1$ is set to $5e^{-3}$, $1$ and $\gamma$ is set to 3.

\subsection{Training and Inference Procedures}
\paragraph{Training} The final training loss terms consist of the following parts: 1) sample reconstruction loss $\mathcal{L}_{\theta}^{MAE}$; 2) structural similarity index (SSIM) loss $\mathcal{L}_{\theta}^{SSIM}$; 3) reconstruction loss for pitch and duration predictor $\mathcal{L}_p$, $\mathcal{L}_{d}$; 4) classification loss for stutter predictor $\mathcal{L}_{BCE}$, $\mathcal{L}_{Focal}$. In the training stage, we randomly select 80\% phonemes spans and mask their corresponding frames since 80\% masking rate shows good performances on both seen and unseen cases. Then we add the stutter embedding to the context condition. The objective functions only take the masked region into consideration.

\paragraph{Inference for reading-style speech editing} Given a speech spectrogram $\boldsymbol{x}$, its original phonemes $\Tilde{{\boldsymbol{p}}}$ and the target phonemes $\boldsymbol{p}$. Denote the spectrogram region that needs to be modified as $\boldsymbol{\mu}$. When the speech recording is reading-style, we do not utilize the stutter predictor. We first use an external alignment tool\footnote{\url{https://github.com/MontrealCorpusTools/Montreal-Forced-Aligner}} to extract the spectrogram-to-phoneme alignments. $\hat{\boldsymbol{x}}$ is the spectrogram masked according to the region $\boldsymbol{\mu}$. FluentSpeech takes $\boldsymbol{p}$, $\hat{\boldsymbol{x}}$, $\boldsymbol{x}$, $\boldsymbol{e}_{spk}$, $\hat{\boldsymbol{e}}_{dur}$, and $\hat{\boldsymbol{e}}_{pitch}$ as inputs and generates the spectrogram of the masked region $\boldsymbol{\mu}$. Finally, we use a pre-trained vocoder to transform this spectrogram into the waveform. 
\paragraph{Inference for stutter removal} When the speech recording is spontaneous, the stutter predictor first predicts the stutter region $\boldsymbol{\mu}^{\prime}$. Since the stutter region $\boldsymbol{\mu}^{\prime}$ also influences the prosody (e.g., duration and pitch) of the neighboring words, we find all of the phoneme spans that overlap with or are adjacent\footnote{The adjacent region is the top-1 adjacent word boundary given by the alignment tool.} to $\boldsymbol{\mu}^{\prime}$ and denote them as $\hat{\boldsymbol{\mu}}$. Then the spectrogram region that needs to be modified can be defined as $\boldsymbol{\mu} = \boldsymbol{\mu}^{\prime} \cup \hat{\boldsymbol{\mu}}$. To make the spontaneous speech fluent, the stutter embedding is not added to the hidden sequence. Following the masked spectrogram reconstruction process in the inference for reading-style speech editing, FluentSpeech is able to perform automatic stutter removal.

\begin{table*}[ht]
\centering
\small
\begin{tabular}{@{}l|ccc|ccc|c@{}}
\toprule
\bfseries \multirow{2}{*}{Method} 
& \multicolumn{3}{c|}{\bfseries VCTK}
& \multicolumn{3}{c|}{\bfseries LibriTTS} & \multirow{2}{*}{\bfseries \#Params.} \\ 
& \bfseries MCD ($\downarrow$) & \bfseries STOI ($\uparrow$) & \bfseries PESQ ($\uparrow$)
& \bfseries MCD ($\downarrow$) & \bfseries STOI ($\uparrow$) & \bfseries PESQ ($\uparrow$) & \\ 
\midrule
EditSpeech     &6.92&0.69&1.43&5.33&0.68&1.35&48.1M   \\
CampNet        &7.83&0.54&1.38&6.51&0.40&1.28&\bfseries14.7M    \\
A${}^3$T       &6.25&0.41&1.18&5.69&0.70&1.39&67.7M   \\
\bfseries FluentSpeech &\bfseries5.86&\bfseries0.81&\bfseries1.91&\bfseries4.74&\bfseries0.78&\bfseries1.82&23.9M \\ 
\bottomrule
\end{tabular}
\caption{The objective audio quality comparisons. We only measure the MCD, STOI, and PESQ of the masked region. MCD and PESQ indicate speech quality, and STOI reflects speech intelligibility}
\label{table_1}
\end{table*}

\section{Experiments}
\subsection{Datasets}
\paragraph{Reading-Style} We evaluate FluentSpeech on two reading-style datasets, including: 1) VCTK~\citep{yamagishi2019vctk}, an English speech corpus uttered by 110 English speakers with various accents; 2) LibriTTS~\citep{zen2019libritts}, a large-scale multi-speaker English corpus of approximately 585 hours of speech. We evaluate the text-based speech editing performance of FluentSpeech and various baselines on these datasets.
\paragraph{Spontaneous} We also evaluate FluentSpeech on the stutter-oriented automatic speech editing (SASE) dataset collected and annotated by us (See Appendix~\ref{appendix: sase dataset} for further details). The SASE dataset consists of approximately 40 hours of spontaneous speech recordings from 46 speakers with various accents. All the audio files are collected from online lectures and courses with accurate official transcripts. Each recording is sampled at 22050 Hz with 16-bit quantization. We evaluate the SASE performance of FluentSpeech and various baselines on this dataset.

For each of the three datasets, we randomly sample 400 samples for testing. We randomly choose 50 samples in the test set for subjective evaluations and use all testing samples for objective evaluations. The ground truth mel-spectrograms are generated from the raw waveform with the frame size 1024 and the hop size 256.

\subsection{Experimental Setup}
\label{Experimental Setup}
\paragraph{Model Configuration} FluentSpeech consists of a linguistic encoder, an acoustic encoder, a masked variance adaptor, a spectrogram denoiser, and a stutter predictor. The linguistic and acoustic encoders consist of multiple feed-forward Transformer blocks~\citep{ren2019fastspeech} with relative position encoding~\citep{shaw2018self} following Glow-TTS~\citep{kim2020glow}. The hidden channel is set to 256. In the spectrogram denoiser, we set $N$ = 20 to stack 20 layers of convolution with the kernel size 3, and we set the dilated factor to 1 (without dilation) at each layer following~\citep{huang2022prodiff}. The number of diffusion steps $T$ is set to 8. The stutter predictor is based on the non-causal WaveNet~\citep{oord2016wavenet} architecture. We have attached more detailed information on the model configuration in Appendix~\ref{appendix:hyper_params}.

\paragraph{Training and Evaluation} We train the FluentSpeech with $T$ = 8 diffusion steps. The FluentSpeech model has been trained for 300,000 steps using 1 NVIDIA 3080 GPU with a batch size of 30 sentences. The adam optimizer is used with $\beta_{1}$ = 0.9, $\beta_{2}$ = 0.98, $\epsilon$ = $10^{-9}$. We utilize HiFi-GAN~\citep{kong2020hifi} (V1) as the vocoder to synthesize waveform from the generated mel-spectrogram in all our experiments. To measure the perceptual quality, we conduct human evaluations with MOS (mean opinion score), CMOS (comparative mean
opinion score), and average preference score on the testing set via Amazon Mechanical Turk (See Appendix~\ref{appendix:subjective_evaluation} for more details). We keep the text content and text modifications consistent among different models to exclude other interference factors, only examining the audio quality. We further measure the objective evaluation metrics, such as MCD~\citep{kubichek1993mel}, STOI~\citep{taal2010short}, and PESQ~\citep{rix2001perceptual}. More information on evaluation has been attached in Appendix~\ref{appendix:objective_evaluation}.

\subsection{Results of Reading-Style Speech Editing}  
\label{evaluation_SE}
We compare the quality of generated audio samples of our FluentSpeech with other baseline systems, including 1) EditSpeech~\citep{tan2021editspeech}; 2) CampNet~\citep{wang2022context}; 3) A${}^3$T~\citep{bai20223} (detailed descriptions can be found in Appendix~\ref{sec:appendix_baseline}). For objective evaluation, we conduct the spectrogram reconstruction experiment to evaluate these systems. As shown in Table \ref{table_1},  FluentSpeech demonstrates superior performance in MCD, PESQ, and STOI metrics.

For subjective evaluation, we manually define modification operations (i.e., insertion, replacement, and deletion) of 50 audio samples. We then conduct the experiments on the VCTK dataset. For each audio sample, we ask at least 10 English speakers to evaluate the generated audios' speech quality and speaker similarity. The results are presented Table~\ref{table_2} and Table~\ref{table_3}. For the seen case, each speaker’s examples would be split into train and test sets. And for the unseen case, the test set contains 10 speakers’ examples, and the other 99 speakers' examples are used for training following~\citep{bai20223}. It can be seen that FluentSpeech achieves the highest perceptual quality and speaker similarity on both seen and unseen settings compared to all baselines, which demonstrates the effectiveness of our proposed context-aware spectrogram denoiser.

\begin{table}[t]
\centering
\small
\begin{tabular}{@{}l|cc@{}}
\toprule
\bfseries Method
& \bfseries Seen
& \bfseries Unseen\\ 
\midrule
EditSpeech     & 4.00 $\pm$ 0.10 & 3.89 $\pm$ 0.09 \\
CampNet        & 3.59 $\pm$ 0.11 & 3.04 $\pm$ 0.18      \\
A${}^3$T       & 4.09 $\pm$ 0.10 & 3.90 $\pm$ 0.10       \\
\bfseries FluentSpeech & \bfseries 4.27 $\pm$ 0.11 & \bfseries 4.18 $\pm$ 0.09\\ 
\bottomrule
\end{tabular}
\caption{The MOS evaluation ($\uparrow$) for speech quality on speech editing task on the VCTK dataset with 95\% confidence intervals.}
\label{table_2}
\end{table}

\begin{table}[t]
\centering
\small
\begin{tabular}{@{}l|cc@{}}
\toprule
\bfseries Method
& \bfseries Seen
& \bfseries Unseen\\ 
\midrule
EditSpeech     & 4.26 $\pm$ 0.10 & 3.90 $\pm$ 0.13\\
CampNet        & 3.93 $\pm$ 0.12 & 3.58 $\pm$ 0.20        \\
A${}^3$T       & 4.27 $\pm$ 0.09 & 3.53 $\pm$ 0.14        \\
\bfseries FluentSpeech & \bfseries 4.42 $\pm$ 0.06 & \bfseries 4.21 $\pm$ 0.11\\ 
\bottomrule
\end{tabular}
\caption{The MOS evaluation ($\uparrow$) for speaker similarity on speech editing task on the VCTK dataset with 95\% confidence intervals.}
\label{table_3}
\end{table}

\begin{figure}[t]
    \centering
    \small
    \includegraphics[width=.98\linewidth]{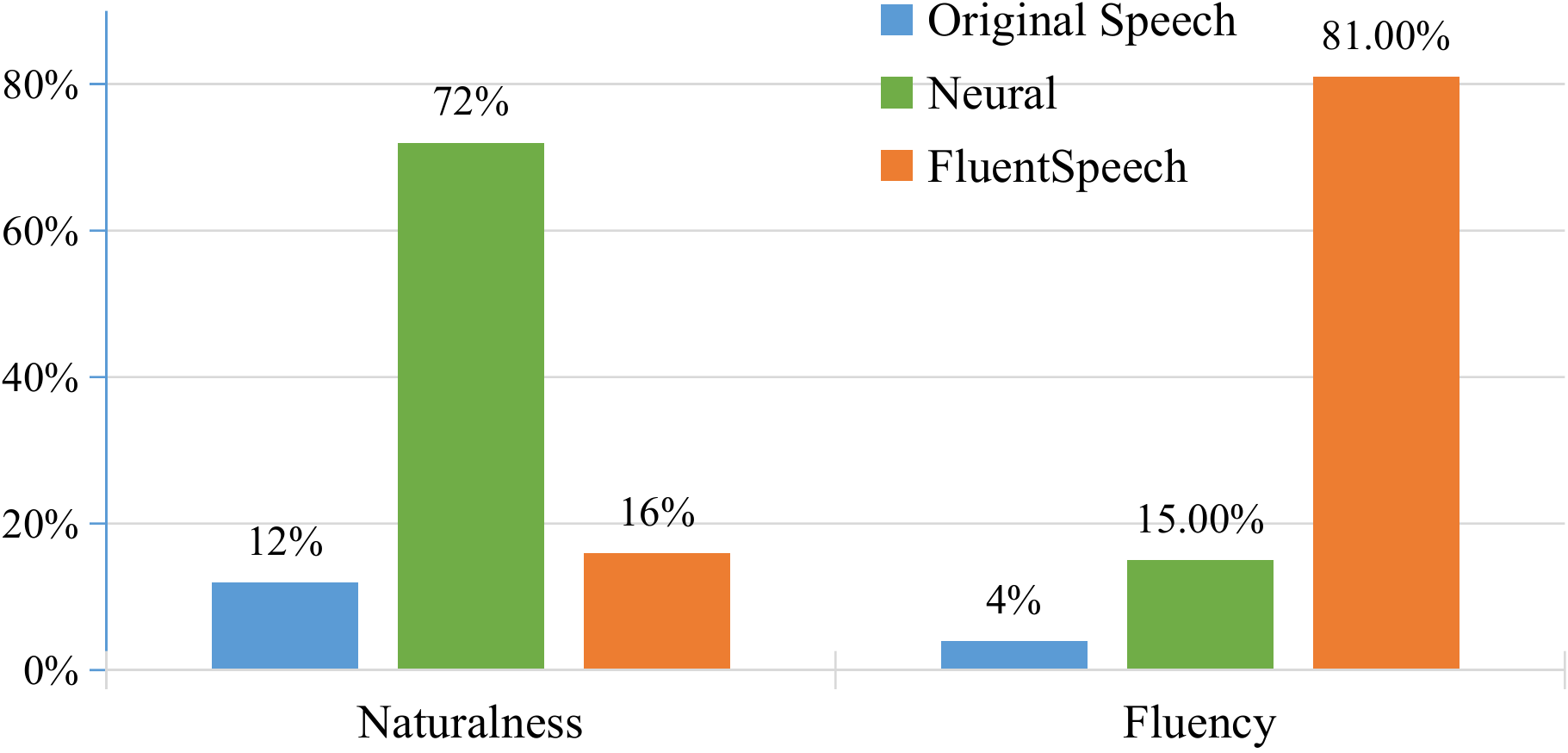}
\caption{Average preference score (\%) evaluation on naturalness and fluency on the SASE dataset, where ``Neural'' stands for ``no preference''.}
\label{stutter_exp_result}
\end{figure}

\begin{table}[t]
\small
\centering
\begin{tabular}{@{}l|ccc@{}}
\toprule
\bfseries Method
& \bfseries Accuracy (\%)
& \bfseries Precision (\%)\\ 
\midrule
FluentSpeech     & 80.5\% & 94.4\% \\
\bottomrule
\end{tabular}
\caption{The stutter localization evaluation ($\uparrow$) on the SASE dataset. Accuracy (\%) denotes the overall accuracy; Precision (\%) indicates the proportion of the correctly classified stutter regions.}
\label{table_4}
\end{table}

\begin{figure*}[!ht]
	\centering
	\begin{minipage}{0.32\linewidth}
		\centering
		\includegraphics[width=1\linewidth,height=80px]{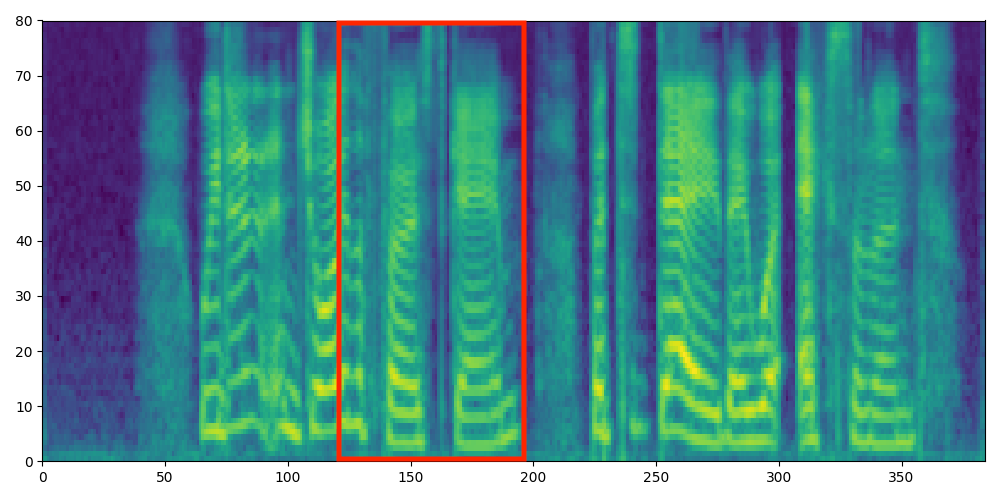}
        \vspace{-6mm}
		\caption*{(a) Ground-truth spectrogram}
	\end{minipage}
	\centering
	\begin{minipage}{0.32\linewidth}
		\centering
		\includegraphics[width=1\linewidth,height=80px]{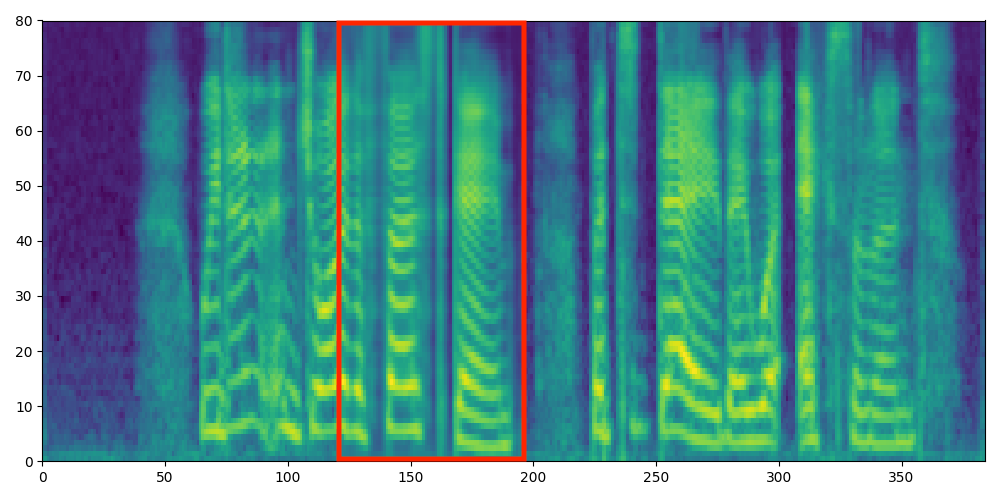}
        \vspace{-6mm}
		\caption*{(b) FluentSpeech}
	\end{minipage}
	\centering
	\begin{minipage}{0.32\linewidth}
		\centering
		\includegraphics[width=1\linewidth,height=80px]{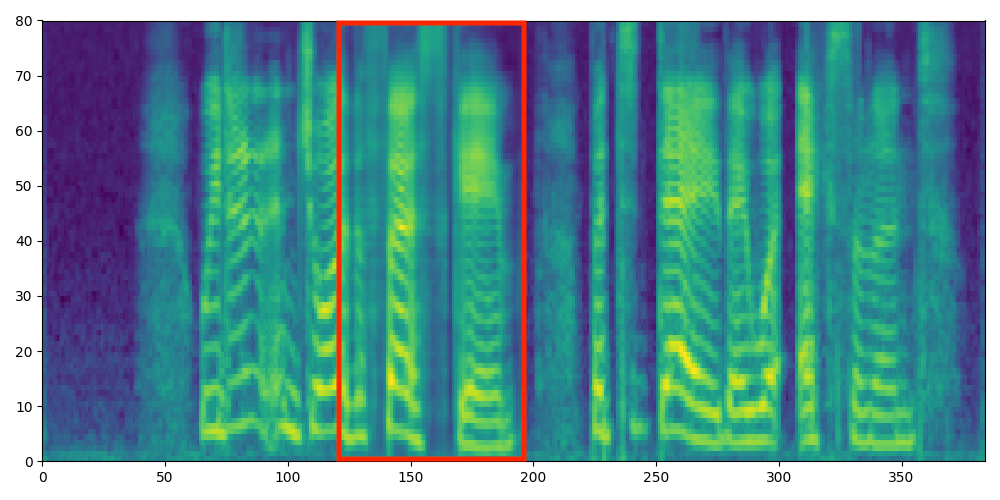}
        \vspace{-6mm}
		\caption*{(c) FluentSpeech wo MDP}
      \label{vis_c}
	\end{minipage}
	\centering
	\begin{minipage}{0.32\linewidth}
		\centering
		\includegraphics[width=1\linewidth,height=80px]{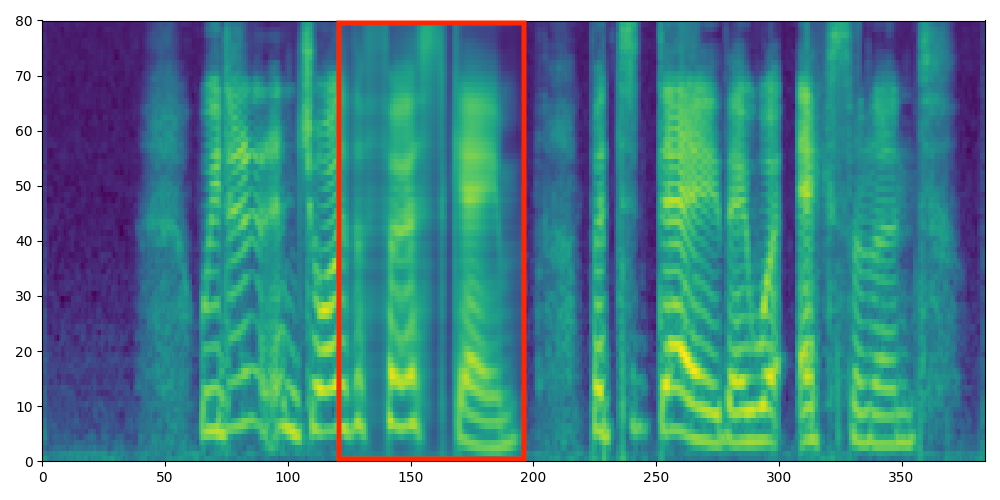}
        \vspace{-6mm}
		\caption*{(d) A3t}
	\end{minipage}
	\centering
	\begin{minipage}{0.32\linewidth}
		\centering
		\includegraphics[width=1\linewidth,height=80px]{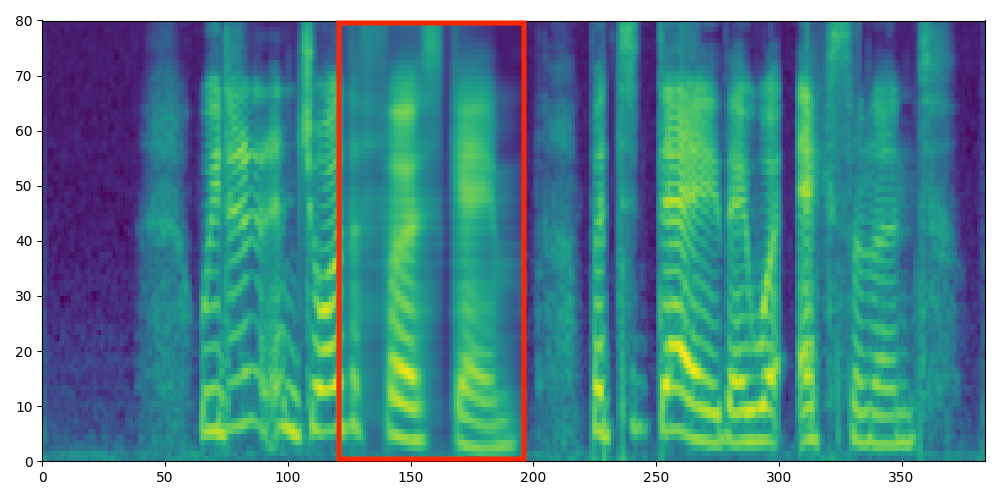}
        \vspace{-6mm}
		\caption*{(e) CampNet}
	\end{minipage}
	\centering
	\begin{minipage}{0.32\linewidth}
		\centering
		\includegraphics[width=1\linewidth,height=80px]{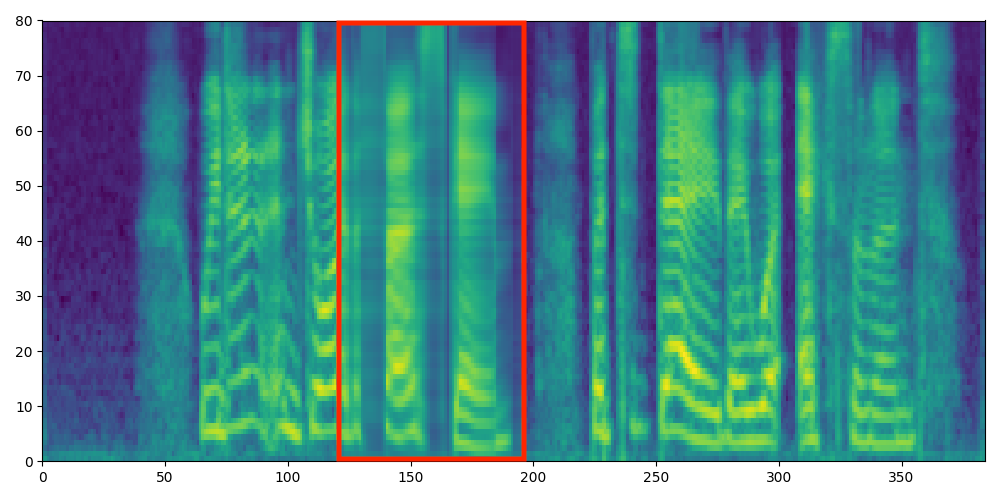}
        \vspace{-6mm}
		\caption*{(f) EditSpeech}
	\end{minipage}
	\centering
	\caption{Visualizations of the ground-truth and generated mel-spectrograms by different speech editing models. Original text is ``We didnt enjoy \textcolor{red}{the first game}, but today they were excellent''. In (b,c,d,e,f) subfigures, the portion with red box is ``\textcolor{red}{the first game}'' which is masked and reconstructed. MDP denotes the masked duration predictor.}
	\label{vis_spec_1}
\end{figure*}

\subsection{Results of Stutter-Oriented Automatic Speech Editing}
We evaluate the accuracy of FluentSpeech on the stutter localization task, and the results are shown in Table~\ref{table_4}. It can be seen that our FluentSpeech achieves 80.5\% accuracy and 94.4\% precision on the stutter localization task.
We then compare the naturalness and fluency of generated audio samples of our FluentSpeech with the original spontaneous recordings. We conduct a subjective average preference score evaluation, where 50 sentences are randomly selected from the test set of our SASE dataset. The listeners are asked to judge which utterance in each pair has better naturalness (or fluency) or no preference in the edited area. As shown in Figure~\ref{stutter_exp_result}, FluentSpeech achieves similar naturalness compared to the original audio. Moreover, the fluency of the speeches generated by our FluentSpeech is significantly improved, which further shows the effectiveness of our stutter-oriented automatic speech editing strategy.

\subsection{Visualizations}
As illustrated in Figure~\ref{vis_spec_1}, we visualize the mel-spectrograms generated by FluentSpeech and baseline systems. We can see that FluentSpeech can generate mel-spectrograms with richer frequency details compared with other baselines, resulting in natural and expressive sounds. Moreover, when we substitute the masked duration predictor with the duration predictor utilized in~\citet{tan2021editspeech,wang2022context,bai20223}, an unnatural transition has occurred in the left boundary of the edited region of FluentSpeech, which demonstrates the effectiveness of our proposed masked duration predictor.

\begin{table}[t]
\label{table_5}
\small
\centering
\begin{tabular}{@{}l|ccc@{}}
\toprule
\bfseries Method
& \bfseries C-MOS
& \bfseries MCD ($\downarrow$)\\ 
\midrule
FluentSpeech     & 0.00 & 4.54 \\
- Stutter Embedding & -0.52 & 4.63 \\
- MDP - MPP + DP + PP & -0.35 & 5.75 \\
- MDP - MPP + DP & -0.24 & 5.15 \\
\bottomrule
\end{tabular}
\caption{Audio quality comparisons on the SASE dataset for ablation study. MDP denotes the masked duration predictor; MPP denotes the masked pitch predictor; DP denotes the duration predictor used in~\citet{tan2021editspeech,bai20223} and PP denotes the pitch predictor proposed in~\citep{ren2020fastspeech}.}
\label{ablation}
\end{table}

\subsection{Ablation Studies}
We conduct ablation studies to demonstrate the effectiveness of several designs in FluentSpeech, including the
stutter embedding and the masked predictors. We perform CMOS and MCD evaluations for these ablation studies. The results are shown in Table~\ref{ablation}. We can see that CMOS drops rapidly when we remove the stutter embedding, indicating that the noise introduced by the text-speech pair's discrepancy greatly reduces the naturalness of the generated audio. Thus, the stutter embedding successfully improves the robustness of our FluentSpeech; Moreover, when we remove the MDP, MPP and use the DP following recent speech editing algorithms~\citep{tan2021editspeech,wang2022context,bai20223}, the speech quality also drops significantly, demonstrating the effectiveness of our proposed masked predictors. It is worth mentioning that the pitch predictor without masked training also results in a performance drop in terms of voice quality.

\section{Conclusion}
In this work, we proposed FluentSpeech, a stutter-oriented automatic speech editing model for stutter removal. FluentSpeech adopts a context-aware spectrogram denoiser to generate high-quality and expressive speeches with rich frequency details. To improve the robustness against stuttering speeches and perform automatic stutter removal, we propose a conditional stutter predictor that localizes the stutter region and injects the stutter embedding into the text hidden sequence to reduce the discrepancy between text and stuttering speech recording. We also contribute a novel stutter-oriented automatic speech editing dataset named SASE, which contains spontaneous speech recordings with time-aligned stutter labels. Experimental results demonstrate that FluentSpeech achieves state-of-the-art performance on speech editing for reading-style speeches. Moreover, FluentSpeech is robust against stuttering speech and demonstrates the ability to improve the fluency of stuttering speech significantly. To the best of our knowledge, FluentSpeech is the first stutter-oriented automatic speech editing model that solves the automatic stutter removal task. Our extensive ablation studies demonstrated that each design in FluentSpeech is effective. We hope that our work will serve as a basis for future stutter-oriented speech editing studies.

\section{Limitations}
We list the limitations of our work as follows. Firstly, the model architecture we use to localize the stuttering speech is simple. Future works could explore a more effective model to perform automatic stutter removal with the help of our SASE dataset. Secondly, we only test the English datasets. And other languages except for English and multi-language stutter-oriented speech editing remain for future works. Finally, after being pre-trained on our SASE dataset, the stutter embedding in FluentSpeech could also be used to inject stutters into the reading-style speech to change its speaking style, and we leave it for future works.

\section{Ethics Statement}
FluentSpeech improves the naturalness of edited speech and promotes the automatic stutter removal of stuttered speech, which may cause unemployment for people with related occupations. Besides, the free manipulation of speeches may bring potential social damage. Further efforts in automatic speaker verification should be made to lower the aforementioned risks.

\section{Acknowledgments}
\label{acknowledgments}
This work was supported in part by the National Key R\&D Program of China under Grant No.2022ZD0162000, National Natural Science Foundation of China under Grant No. 62222211, Grant No.61836002 and Grant No.62072397.

\bibliography{anthology,custom}
\bibliographystyle{acl_natbib}

\appendix

\section{Detailed Experimental Settings}
\label{appendix:details_of_experiments}
\subsection{Model Configurations}
\label{appendix:hyper_params}
We list the model hyper-parameters of FluentSpeech in Table~\ref{tab:hyperparameters_ps}.

\begin{table*}[h]
\small
\centering
\begin{tabular}{l|c|c|c}
\toprule
\multicolumn{2}{c|}{Hyperparameter}   & FluentSpeech & Number of parameters \\ 
\midrule
\multirow{5}{*}{Text Encoder} 
&Phoneme Embedding           &192 & \multirow{5}{*}{3.7M}  \\
&Encoder Layers              &4   \\
&Encoder Hidden              &192     \\                     
&Encoder Conv1d Kernel       &5   \\    
&Encoder Conv1D Filter Size  &384\\                          
\midrule
\multirow{3}{*}{Context Condition}         
& Predictor Conv1D Kernel        & 3 & \multirow{3}{*}{5.8M}\\    
& Predictor Conv1D Filter Size   & 256  \\    
& Predictor Dropout              & 0.4  \\ 
\midrule
\multirow{5}{*}{Spectrogram Denoiser}     
&Diffusion Embedding                &  256 & \multirow{5}{*}{14.4M}  \\   
&Residual Layers                    &  20 \\       
&Residual Channels                  &  256 \\     
&WaveNet Conv1d Kernel              &  3 \\       
&WaveNet Conv1d Filter              &  512 \\ 
\midrule
\multicolumn{2}{c|}{Total Number of Parameters}   & \multicolumn{2}{c|}{23.9M}  \\
\bottomrule
\end{tabular}
\caption{Hyperparameters of FluentSpeech models.}
\label{tab:hyperparameters_ps}
\end{table*}

\subsection{Details of Baseline Systems}
\label{sec:appendix_baseline}
EditSpeech~\citep{tan2021editspeech} is a speech-editing system that introduces partial inference and bidirectional fusion to sequence-to-sequence neural TTS model. EditSpeech trains two conventional autoregressive TTS models, one left-to-right and the other right-to-left. For decoding, the left-to-right TTS model and the right-to-left TTS model generate the modified region simultaneously. Finally, the two synthesized speeches are fused for the final output.
CampNet~\citep{wang2022context} propose a context-aware mask prediction network (CampNet) to simulate the process of text-based speech editing. Three text-based speech editing operations based on CampNet are designed: deletion, replacement, and insertion. And a word-level autoregressive generation method is proposed to improve the editing length.
A${}^3$T~\citep{bai20223} propose the alignment-aware acoustic-text pre-training, a BERT-style pre-training model, which takes both phonemes and partially-masked spectrograms as inputs. The alignment embedding from external alignment tools is introduced into the Conformer-based~\citep{gulati2020conformer,guo2021recent} backbone to improve the speech quality.

\begin{figure*}[!ht]
    \centering
    \small
    \subfigure[Screenshot of MOS testing for speech quality in the speech editing evaluation.]{
        \centering
        \includegraphics[width=.85\linewidth]{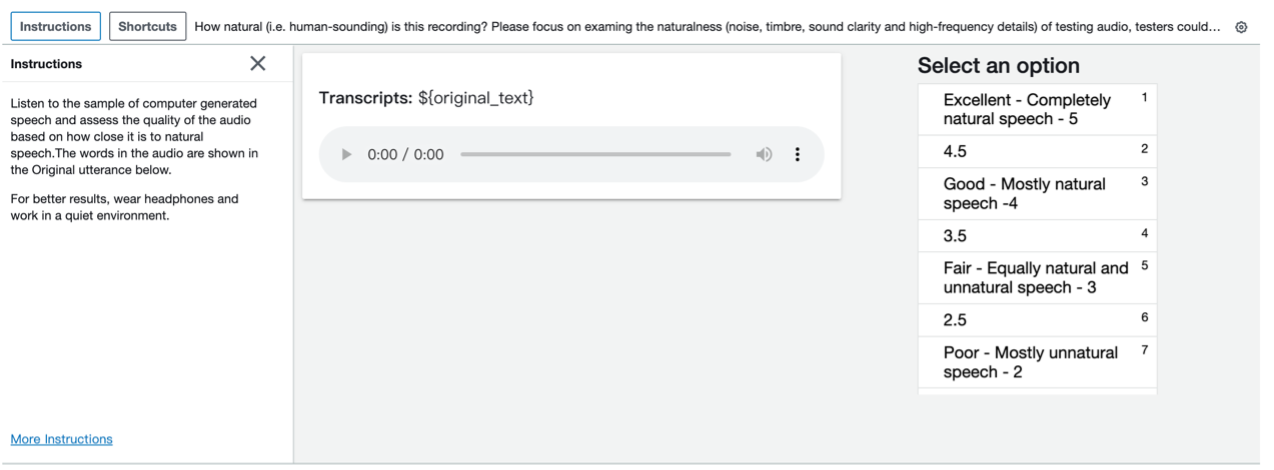}
        \label{MOS_quality}
    }
    \subfigure[Screenshot of MOS testing for speaker similarity in the speech editing evaluation.]{
        \centering
        \includegraphics[width=.85\linewidth]{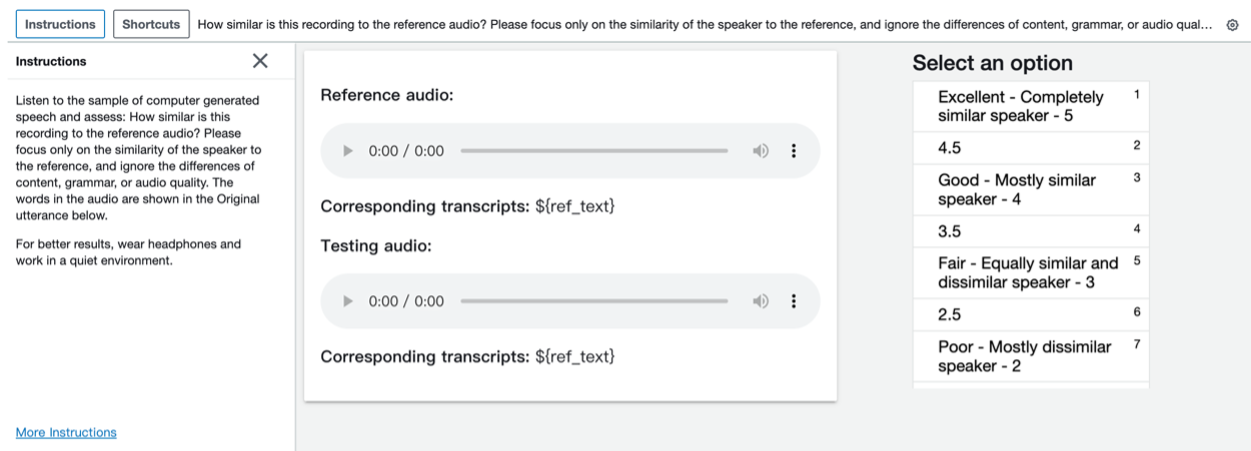}
        \label{MOS_similarity}
    }
    \subfigure[Screenshot of average preference score testing for speech quality in the stutter removal evaluation.]{
        \centering
        \includegraphics[width=.85\linewidth]{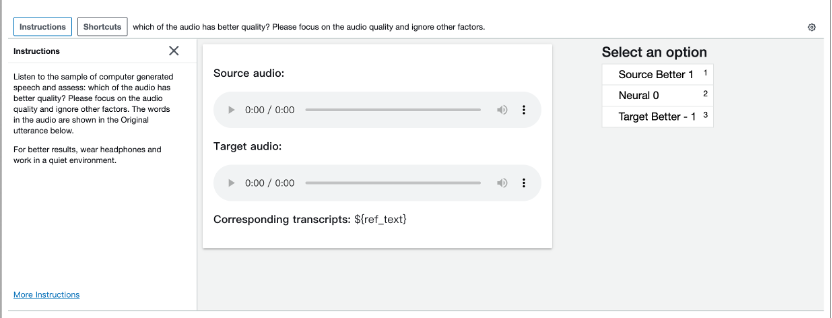}
        \label{MOSQ}
    }
    \subfigure[Screenshot of average preference score testing for speech fluency in the stutter removal evaluation.]{
        \centering
        \includegraphics[width=.85\linewidth]{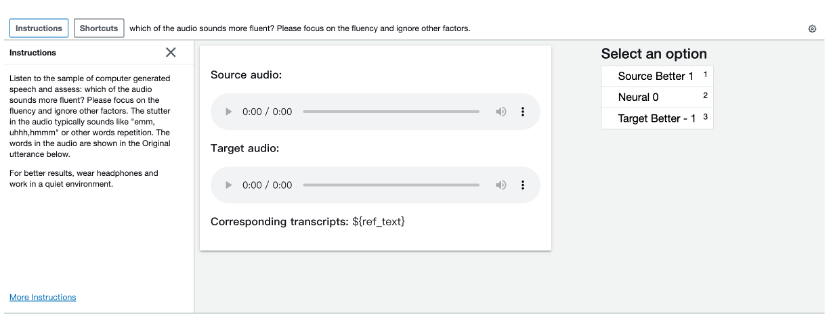}
        \label{MOSF}
    }
\caption{Screenshots of subjective evaluations.}
\label{Screenshots_subjective_evaluations}
\end{figure*}

\subsection{Details in Subjective Evaluation}
\label{appendix:subjective_evaluation}
We perform the subjective evaluation on Amazon Mechanical Turk (MTurk). For speech editing evaluations, we randomly select 50 samples from the test set and manually define modification operations (i.e., insertion, replacement, and deletion) for these audio samples. We use FluentSpeech and the baseline speech editing systems to edit the audio samples. Each generated audio has been listened to by at least 10 native listeners. We paid \$8 to participants hourly and spent about \$400 on participant compensation. We tell the participants that the data will be used in scientific research.

\begin{itemize}
\item For audio quality evaluations (MOS), each tester is asked to evaluate the subjective naturalness of a sentence on a 1-5 Likert scale, and we tell listeners to``\textit{assess the quality of the audio based on how close it is to natural speech}''. 
\item For speaker similarity evaluations (MOS), listeners are asked to compare pairs of audio generated by systems A and ground-truth B and indicate the speaker similarity of the two audio and choose the scores on a 1-5 similar scale. We tell listeners to answer ``\textit{How similar is this recording to the reference audio? Please focus only on the similarity of the speaker to the reference, and ignore the differences of content, grammar, or audio quality}''. The screenshots of instructions for speech editing tests are shown in Figure~\ref{MOS_quality} and Figure~\ref{MOS_similarity}.
\item For stutter removal evaluations, we perform average preference score tests for speech quality and fluency. For the speech quality AB test, each listener is asked to select their preferred audio according to audio quality. We tell listeners to answer ``\textit{Which of the audio has better quality? Please focus on the audio quality and ignore other factors}''. For the speech fluency AB test, each listener is asked to select the audio they prefer according to audio fluency, and we tell listeners to answer ``\textit{Which of the audio sounds more fluent? Please focus on speech fluency and ignore other factors. The stutter in the audio typically sounds like ``emm'', ``uhhh'', ``hmmm'', or words repetition}''. The screenshots of instructions for stutter removal evaluations are shown in Figure~\ref{MOSQ} and Figure~\ref{MOSF}.
\end{itemize}

\subsection{Details in Objective Evaluation}
\label{appendix:objective_evaluation}
The effectiveness of our FluentSpeech is measured by  MCD~\citep{toda2007voice}, STOI~\citep{taal2011algorithm},PESQ~\citep{hu2007evaluation} metrics. MCD measures the Euclidean distance between two mel cepstral sequences, which describes the global spectral characteristics of audio signals. PESQ indicates speech quality, and STOI reflects speech intelligibility~\citep{zhang2018training}. The lower MCD and
higher PESQ, STOI represent better performance in the generated speech. Denote $m^t = [m_{1}^t,\ldots,m_{L}^t]$ and $m^c = [m_{1}^t,\ldots,m_{L}^t]$ as two mel cepstral sequences. The traditional MCD measure is given by:
\begin{equation}
MCD[dB]=\frac{10}{ln10}\sqrt{2\sum_{i=1}^{L}(m_{i}^{t}-m_{i}^{c} )^{2}} \ , 
\label{mcd}
\end{equation}
where $L$ is the order of mel cepstrum and $L$ is 34 in our implementation.

The traditional PESQ measure is given by:
\begin{equation}
PESQ=a_{0} + a_{1}D_{ind} + a_{2}A_{ind} \ ,
\label{pesq}
\end{equation}
where $a_{0}$,$a_{1}$,$a_{2}$ are the parameters, $D_{ind}$ represents the average disturbance value and $A_{ind}$ represents the average asymmetrical disturbance values.

STOI is a function of a TF-dependent intermediate intelligibility measure, which compares the temporal envelopes of the clean and degraded speech in short-time regions by means of a correlation coefficient. The following vector notation is used to denote the short-time temporal envelope of the clean speech:
\begin{equation}
\small
x_{j,m} = [X_{j}(m-N+1),X_{j}(m-N+2),...,X_{j}(m)]^{T} \ ,
\label{stoi}
\end{equation}
where $N=30$ which equals an analysis length of 384 ms.

\begin{table}[!t]
\small
\center
\begin{tabular}{l|c}
\toprule
\textbf{Method}              & \multicolumn{1}{c}{\textbf{Duration Error (ms) ($\downarrow$)}} \\ 
\midrule
DP & 152.9 \\
\textbf{MDP} & \textbf{99.9} \\ 
\bottomrule
\end{tabular}
\caption{Average duration error comparisons on the VCTK dataset. DP denotes the duration predictor used in~\citet{tan2021editspeech,bai20223} and MDP denotes the masked duration predictor in our FluentSpeech.}
\label{duration errors}
\end{table}

\begin{table}[!t]
\small
\center
\begin{tabular}{l|c}
\toprule
\textbf{Method}     & \textbf{Average Pitch Error ($\downarrow$)} \\ \midrule
EditSpeech          & 5571                         \\
CampNet             & 6758                         \\
A${}^3$T            & 4595                         \\
FluentSpeech w/ PP & 4134                         \\
\textbf{FluentSpeech w/ MPP} & \textbf{2276}         \\ \bottomrule
\end{tabular}
\caption{Average pitch error comparisons on the VCTK dataset. PP denotes the pitch predictor proposed in~\citet{ren2020fastspeech} and MPP denotes the masked pitch predictor in our FluentSpeech.}
 \label{pitch errors}
\end{table}

\begin{figure*}[!ht]
    \centering
    \small
    \includegraphics[width=.98\linewidth]{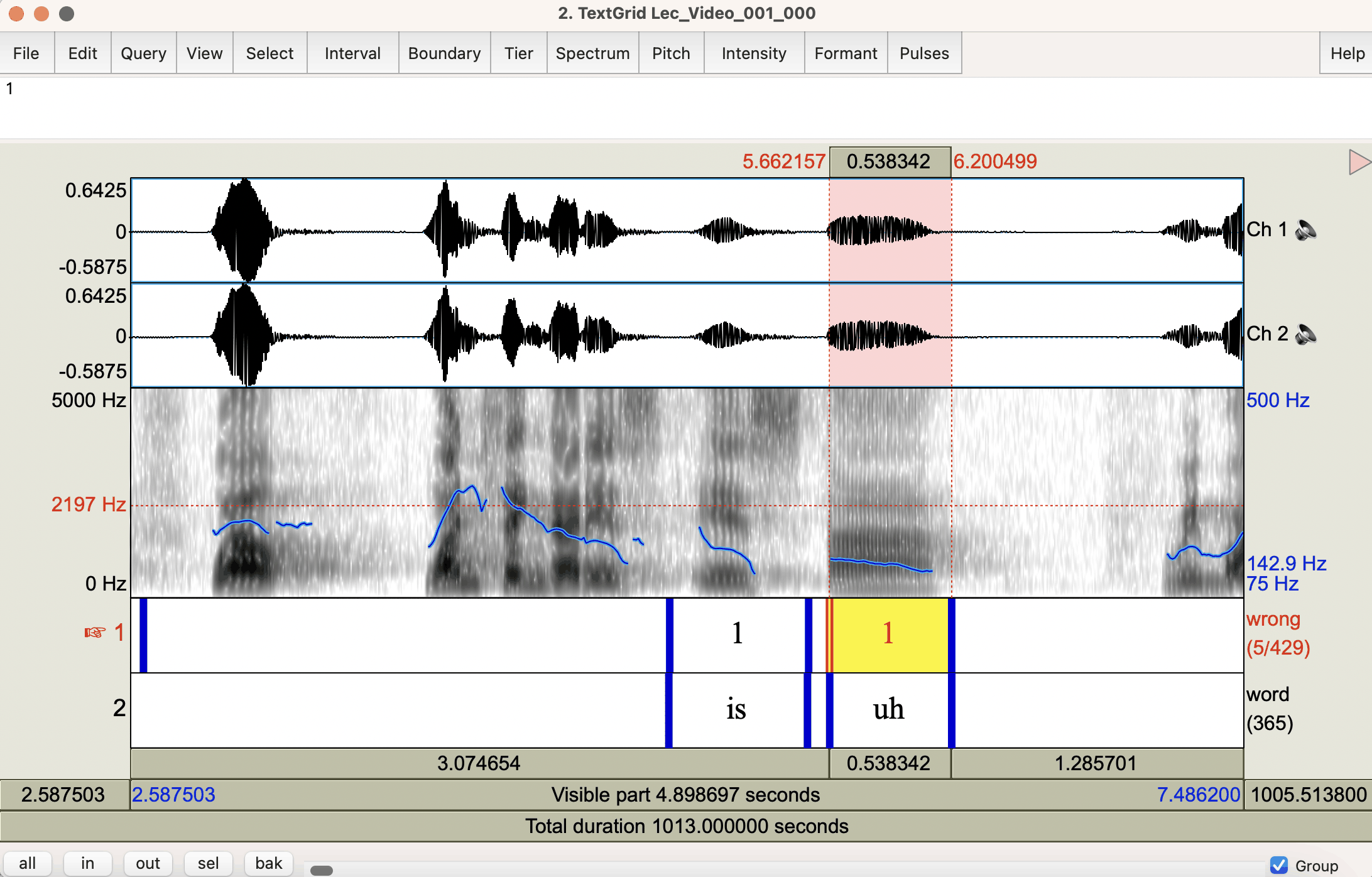}
\caption{Screenshot of our SASE dataset.}
\label{dataset_screenshot}
\end{figure*}

\section{Detailed analysis of duration and pitch}
\label{sec:appendix duration and pitch}
To further dive into the detailed performance of our model, we evaluate the duration and pitch errors between our FluentSpeech and the baseline models. For duration errors, the ground truth duration is obtained from the Montreal Forced Aligner (MFA)~\citep{mcauliffe2017montreal}. We calculate MSE of word-level durations for the duration predictor (DP) used in~\citet{tan2021editspeech,bai20223} and the masked duration predictor (MDP) in FluentSpeech. The results on the VCTK dataset are shown in Table~\ref{duration errors}. It can be seen that the masked duration predictor predicts more accurate duration, demonstrating the effectiveness of the masked prediction training. For pitch errors, we compare our FluentSpeech with all other baseline models. We firstly extract frame-level pitch information using parselmouth\footnote{\url{https://github.com/YannickJadoul/Parselmouth}}, then calculate the MSE of the mean pitch distance between the model-generated speeches and the ground-truth speeches. The results on the VCTK dataset are shown in table~\ref{pitch errors}. It can be seen that FluentSpeech achieves the lowest average pitch error. Moreover, the average pitch error of FluentSpeech with the masked pitch predictor (MPP) is significantly lower than the FluentSpeech with the pitch predictor proposed in~\citet{ren2020fastspeech}, demonstrating the effectiveness of our masked pitch predictor.

\section{More details of SASE dataset}
\label{appendix: sase dataset}
The SASE dataset consists of approximately 40 hours of spontaneous speech recordings from 46 speakers with various accents. The speech recordings are crawled from online lectures and courses with accurate official transcripts. Each recording is sampled at 22050 Hz with 16-bit quantization. We substitute the speakers' names with speaker IDs to protect their personal information, and the dataset can only be accessed for research purposes. 

To obtain the time-aligned stutter labels, we recruit annotators from a crowdsourcing platform, Zhengshu Technology, to label the stuttering region according to the audio and transcription. Specifically, the stuttering region may be 1) stammers and repetitive words, for instance, ``I am go...go...going... out for a...a...a... trip''; 2) filled pauses (FP) such as ``em, um, then, due to, uh, the speaker’s custom of speaking''; 3) sudden occasions such as cough, voice crack, etc. The annotators are asked to mark the corresponding time boundaries and give the stuttering label as shown in Figure~\ref{dataset_screenshot}. We then use the given timestamps in the official transcriptions to cut the audio and text into fragments ranging from 7 to 10 seconds. Finally, we convert each text sequence into phoneme sequence with an open-source grapheme-to-phoneme tool\footnote{\url{https://github.com/Kyubyong/g2p}}. The audio samples in our SASE dataset are available at \url{https://speechai-demo.github.io/FluentSpeech/}.

\end{document}